\documentclass[sn-vancouver,iicol,Numbered]{sn-jnl} 

\usepackage{graphicx}%
\usepackage{multirow}%
\usepackage{amsmath,amssymb,amsfonts}%
\usepackage{amsthm}%
\usepackage{mathrsfs}%
\usepackage[title]{appendix}%
\usepackage{xcolor}%
\usepackage{textcomp}%
\usepackage{manyfoot}%
\usepackage{booktabs}%
\usepackage{algorithm}%
\usepackage{algorithmicx}%
\usepackage{algpseudocode}%
\usepackage{listings}%
\usepackage{float}
\usepackage{tikz}
\usepackage{marginnote}

\usepackage{mathtools}
\usepackage{setspace}
\usepackage{xspace}
\usepackage{soul}
\usepackage{color, colortbl}

\usepackage{graphicx} 
\usepackage{bm} 
\usepackage{array} 
\usepackage{caption} 
\DeclareMathOperator*{\argmin}{arg\,min}

\newcommand{\Complex}{{\mathbb{C}}}

\renewcommand{\eqref}[1]{Equation~(\ref{eq:#1})}

\newcommand{\tabref}[1]{Table~\ref{tab:#1}}
\newcommand{\figref}[1]{Figure~\ref{fig:#1}}
\newcommand{\alref}[1]{Algorithm~\ref{alg:#1}}
\newcommand{\linref}[1]{Line~\ref{lin:#1}}

\newcommand{\textr}[1]{\textcolor{black}{#1}}
\newcommand{\textb}[1]{\textcolor{black}{#1}}

\renewcommand{\tilde}{\widetilde}

\newcommand{\defn}{\triangleq}

\newcommand{\tvec}[1]{\ensuremath{\Tilde{\boldsymbol{#1}}}}

\newcommand{\hvec}[1]{\ensuremath{\Hat{\boldsymbol{#1}}}}
\renewcommand{\vec}[1]{\ensuremath{\boldsymbol{#1}}}

\newcommand{\herm}{^\textsf{H}}
\newcommand{\reside}{ReSiDe\xspace}
\newcommand{\resides}{ReSiDe-S\xspace}
\newcommand{\residem}{ReSiDe-M\xspace}
\definecolor{Gray}{gray}{0.5}
\definecolor{Light_Gray}{gray}{0.7}

\newcolumntype{C}[1]{>{\centering\arraybackslash}p{#1}}

\theoremstyle{thmstyleone}%
\theoremstyle{thmstyletwo}%

\theoremstyle{thmstylethree}%

\begin{document}

\title[Article Title]{MRI Recovery with Self-Calibrated Denoisers without Fully-Sampled Data}


\author[1,2]{\fnm{Muhammad} \sur{Shafique}}\email{shafique.9@osu.edu}
\equalcont{These authors contributed equally to this work.}

\author[1]{\fnm{Sizhuo} \sur{Liu}}\email{liu.6221@osu.edu}
\equalcont{These authors contributed equally to this work.}

\author[3]{\fnm{Philip} \sur{Schniter}}\email{schniter.1@osu.edu}

\author*[1]{\fnm{Rizwan} \sur{Ahmad}}\email{ahmad.46@osu.edu}

\affil[1]{\orgdiv{Biomedical Engineering}, \orgname{Ohio State University}, \orgaddress{\city{Columbus}, \state{OH} \postcode{43210}, \country{USA}}}

\affil[2]{\orgdiv{Electrical and Computer Engineering}, \orgname{COMSATS University}, \orgaddress{\city{Islamabad}, \country{Pakistan}}}

\affil[3]{\orgdiv{Electrical and Computer Engineering}, \orgname{Ohio State University}, \orgaddress{\city{Columbus}, \state{OH} \postcode{43210}, \country{USA}}}



\abstract{
\textbf{Objective} Acquiring fully sampled training data is challenging for many MRI applications. We present a self-supervised image reconstruction method, termed \reside, capable of recovering images solely from undersampled data.
\vspace{1mm}

\textbf{Materials and Methods} \reside is inspired by plug-and-play (PnP) methods, but unlike traditional PnP approaches that utilize pre-trained denoisers, 
\reside iteratively trains the denoiser on the image or images that are being reconstructed. We introduce two variations of our method: \resides and \residem. \resides is scan-specific and works with a single set of undersampled measurements, while \residem operates on multiple sets of undersampled measurements and provides faster inference.
Studies I, II, and III compare \resides and \residem against other self-supervised or unsupervised methods using data from T1- and T2-weighted brain MRI, MRXCAT digital perfusion phantom, and first-pass cardiac perfusion, respectively. 
\vspace{1mm}

\textbf{Results} \resides and \residem outperform other methods in terms of peak signal-to-noise ratio and structural similarity index measure for Studies I and II, and in terms of expert scoring for Study III.
\vspace{1mm}

\textbf{Discussion} We present a self-supervised image reconstruction method and validate it in both static and dynamic MRI applications. These developments can benefit MRI applications where the availability of fully sampled training data is limited.
}




\maketitle

\section{Introduction}\label{sec:intro}
Magnetic resonance imaging (MRI) is a well-established diagnostic tool that offers several advantages over other imaging modalities, including excellent soft-tissue contrast, high spatial and temporal resolution, multiple contrast mechanisms, and radiation-free acquisition. MRI is routinely used in various clinical applications, including neuro, musculoskeletal, abdominal, and cardiovascular imaging. However, long scan times remain a challenge in MRI, as they can reduce patient comfort, increase sensitivity to motion, and decrease patient throughput. Consequently, accelerating MRI has become a highly active area of research \cite{ravishankar2019image}.



In parallel MRI (pMRI), which is available on all commercial MRI scanners, data are acquired simultaneously across multiple receive coils \cite{pruessmann1999sense}. 
\textr{Typically, pMRI can speed up the acquisition process by a factor of two to three for 2D planar imaging, while higher acceleration factors are achievable in multi-band and 3D imaging.} To achieve further acceleration, pMRI can be combined with methods that utilize prior information about the image. For instance, compressed sensing (CS) leverages sparsity-based priors and can be effectively paired with pMRI \cite{lustig2007sparse}. The combination of pMRI and CS achieves higher acceleration rates than pMRI alone, and these reconstruction methods are increasingly available on commercial scanners \cite{forman2016compressed}. 
More recently, deep learning (DL) methods have been developed to reconstruct images from highly undersampled MRI data. Several studies indicate that DL methods can outperform sparsity-driven compressed sensing (CS) methods in terms of image quality \cite{zbontar2018fastmri,aggarwal2018modl}. \textr{In particular, end-to-end variational networks (VarNet), as early DL-based extensions of compressed sensing, have shown strong performance across various MRI reconstruction tasks and are considered a competitive benchmark in the field \cite{sriram2020end}. Most DL methods, including VarNet,} are based on supervised learning, where a reconstruction network is trained on a large, fully sampled training dataset \cite{wang2021deep}. Outside of a few 2D applications, such training datasets are often unavailable. For other applications, such as cardiac imaging, collecting \textr{full resolution fully sampled} data may not be feasible \cite{chen2020ocmr}. Consequently, self-supervised DL (SSDL) techniques have recently gained significant interest for MRI reconstruction \cite{zeng2021review}. These techniques do not require fully sampled training datasets and instead utilize the redundant information within the undersampled data to guide the training process. 

Several SSDL methods have been proposed for image denoising, including single-instance deep generative prior methods such as deep image prior (DIP) and deep decoder \cite{ulyanov2018deep,heckel2018deep}. These methods model an image as the output of a generator network, with both network parameters and input code vectors trained on an image-specific basis. Another popular SSDL denoising method is Noise2Noise \cite{lehtinen2018noise2noise}, which denoises images using two noisy copies of the same image. However, acquiring multiple copies of an image is not practical for most MRI applications. To address this issue, other SSDL-based denoising methods have recently been proposed, including Noise2Void  \cite{krull2019noise2void}, Noise2Self \cite{batson2019noise2self}, and Self2Self \cite{quan2020self2self}, which operate on a single noisy image. These methods train a network to predict a pixel from its neighboring pixels or predict one group of pixels from another. Since noise is assumed to be independent across pixels, the network denoises the image by implicitly learning the underlying structure in the image. For SSDL-based denoising, Xu et al. took a different approach and proposed a method called Noisy-As-Clean \cite{xu2020noisy}. This method works by adding synthetic noise to the noisy images and training a denoising network to remove the added noise. The trained network is then used to denoise the images on which it was trained.  In a separate development, Stein’s unbiased risk estimator (SURE)-based loss was used for unsupervised training of denoising networks  \cite{zhussip2019extending}.

The application of SSDL extends beyond image denoising, and many of these methods can be applied to solve inverse problems, including image reconstruction. For instance, DIP can readily solve inverse problems with a known forward operator. Yoo et al. employed DIP for reconstructing dynamic MRI \cite{yoo2021time}. More recently, Bell et al. introduced a more robust adaptation of DIP by training the network to function as a denoiser instead of a generator \cite{bell2023robust}. In a separate study, Hamilton et al. integrated low-rank constraint with DIP and applied it to free-breathing cardiac imaging \cite{hamilton2023low}. Scan-specific robust artificial neural network for k-space interpolation (RAKI) is another self-supervised method proposed for MRI reconstruction \cite{akccakaya2019scan}. RAKI is similar to Noise2Void but operates in k-space. In RAKI, a network is first trained on the fully sampled auto-calibration signal (ACS) region and then used to predict missing k-space samples from neighboring measured samples. Both RAKI and its recent extension, called residual RAKI \cite{zhang2022residual}, can be viewed as nonlinear extensions of traditional GRAPPA \cite{griswold2002generalized}. However, due to their scan-specific nature, DIP and RAKI are computationally slow, which limits their translational potential. In 2020, Yaman et al. proposed self-supervised learning via data undersampled (SSDU), a self-supervised learning method that resembles Noise2Self, but with a loss function defined in k-space \cite{yaman2020self}. In SSDU, the acquired undersampled k-space is divided into two subsets, and an unrolled network is trained to infer images from the first subset such that those images are consistent with the second subset. At the inference stage, the trained network in SSDU can rapidly map an undersampled, aliased image to a fully sampled image. More recently, Millard and Chiew used the Noisier2Noise \cite{moran2020noisier2noise} framework to further improve SSDU by ensuring that the sampling masks of the two subsets have similar distributions \cite{millard2023theoretical}. Cole et al. \cite{cole2020unsupervised} also proposed training a network to map undersampled, aliased images to fully sampled images but with an adversarial loss, where the discriminator is fed two unrelated undersampled images: one from the image reconstruction network output and one from an independent set of measurements. \textr{Finally, generalized SURE (GSURE), which extends SURE to inverse problems, has been proposed \cite{eldar2009}. However, for practical inverse problems, the projected mean squared error used in GSURE often poorly approximates the true mean squared error. Aggarwal et al. \cite{aggarwal2022ensure} addressed this issue by using an ensemble approach, but it requires a collection of data undersampled with different sampling patterns, which may not be available. In another application of SURE for inverse problems \cite{zhussip2019training}, Zhussip et al. used SURE and an approximate message passing algorithm to train the denoiser from corrupted images for  Gaussian measurement matrices. However, this method relies on the residual at each iteration of image reconstruction being approximately Gaussian, which is not typically the case for non-trivial forward operators.}


In this work, we propose an SSDL method, called recovery with a self-calibrated denoiser (\reside), for image reconstruction. Our main contribution involves integrating the self-supervised denoising scheme of Noisy-As-Clean \cite{xu2020noisy}  with the PnP framework to solve the inverse problem in MRI reconstruction. Additionally, we employ the discrepancy principle to iteratively adapt the strength of the denoiser, providing faster convergence and more control over the quality of the final reconstruction. Finally, we propose a faster, more robust version of \reside that is applicable to cases where multiple undersampled sets of measurements are available. Using data from brain MRI, MRXCAT perfusion phantom, and first-pass perfusion MRI, we show that \reside outperforms other self-supervised and unsupervised image reconstruction methods. These developments significantly expand our preliminary description of \reside \cite{liu2022mri}, which did not include auto-tuning, was applicable only to a single set of measurements, and used only one T1 and one T2-weighted image for validation.

\section{Materials and Methods}\label{sec:the}
\subsection{Plug-and-Play based Recovery}
The MRI reconstruction entails estimating the underlying image from noisy and potentially undersampled k-space measurements. The measured noisy k-space data are related to the image via
\begin{equation}
\label{eq:forward}
\vec{y} = \vec{A}\vec{x} + \vec{\eta},
\end{equation}
where $\vec{x}\in\Complex^N$ is a vectorized $N$-pixel image, $\vec{y}\in\Complex^M$ is the \textr{pre-whitened and scaled pMRI data from $C$ receive coils, $\vec{\eta}\in\Complex^M$ is circularly symmetric zero-mean white Gaussian noise with known variance $\sigma^2$ that depends on the data scaling,} and $\vec{A}\in \Complex^{M\times N}$ is a known forward operator that subsumes coil sensitivity maps, discrete Fourier transform, and k-space undersampling. 

To reduce acquisition time, the k-space data are often prospectively undersampled to achieve an acceleration rate $R\defn \frac{CN}{M} > 1$. At high acceleration rates, the problem in \eqref{forward} becomes ill-posed. In that case, a common remedy is to inject prior knowledge about $\vec{x}$ using a regularizer, resulting in the optimization problem of the form
\begin{equation}
\hvec{x} = \argmin_{\vec{x}} \frac{1}{\sigma^2}\|\vec{y} - \vec{A}\vec{x} \|^2_2 + \mathcal{R}(\vec{x}),
\label{eq:optim}
\end{equation}
where the first term enforces data consistency and $\mathcal{R}(\cdot)$ represents the regularizer. For CS-based MRI reconstruction, popular choices for $\mathcal{R}(\vec{x})$ include total variation \cite{block2007undersampled} and $\lambda \|\vec{\Phi} \vec{x}\|_1$, where $\vec{\Phi}$ represents a linear sparsifying transfrom, and $\lambda > 0$ controls the regularization strength \cite{lustig2007sparse}. It has been shown that a simple sparsity-based regularizer may not fully capture the rich structure in medical images \cite{ahmad2015iteratively}. To leverage more complex priors, Venkatakrishnan et al. proposed an algorithmic framework called plug-and-play (PnP) \cite{venkatakrishnan2013plug}. In PnP, an off-the-shelf image denoiser, $\vec{f}(\cdot)$, is called within an iterative algorithm, e.g., primal-dual splitting (PDS) \cite{ono2017primal}. A PDS-based implementation of PnP is given in \alref{pds}. In the subsequent sections, we will use this algorithm as a starting point to explain \reside. Note, the implementation of PnP or \reside is not limited to PDS and can be carried out using other algorithms \cite{ahmad2020plug}. 

\begin{algorithm*}[ht!]
\onehalfspacing
\caption{PnP implemented using PDS}\label{alg:pds}
\begin{algorithmic}[1]
\Require $\nu>0$, $\sigma$, $\vec{A}$, $\vec{y}$, $\vec{f}$, $\gamma = \frac{\nu}{\sigma^2}\|\vec{A}\|_2^2$, $\vec{x}_0 = \vec{A}{\herm}\vec{y}$, $\vec{z}_0 = \vec{A}\vec{x}_0 - \vec{y}$
\Ensure $\hvec{x}=\vec{x}_T$ \texttt{~~\% final reconstructed image}
\For{$t=1,2,\dots,T$}
    \State{$\vec{u}_t = \vec{x}_{t-1} - \frac{\nu}{\sigma^2}\vec{A}\herm\vec{z}_{t-1}$} \texttt{~~\% intermediate image}
    \label{lin:pds-u}
	\State{$\vec{x}_t = \vec{f}(\vec{u}_t)$} \texttt{~~\% denoising}
    \label{lin:pds-denoise}
	\State{$\vec{z}_t=\frac{\gamma}{1+\gamma}\vec{z}_{t-1} + \frac{1}{1+\gamma}(\vec{A}(2\vec{x}_t - \vec{x}_{t-1}) - \vec{y})$}
\EndFor
\end{algorithmic}
\end{algorithm*}

\subsection{\resides}\label{sec:resides}
\resides is a scan-specific technique that operates on a single set of undersampled measurements, $\vec{y}$. A PDS-based implementation of \resides is given in \alref{resides}. \resides differs from PnP in the way the denoiser $\vec{f}(\cdot)$ is trained. The denoiser used in PnP is either generic or pre-trained using an additional training dataset. In contrast, the denoiser in \resides is progressively trained from the image being recovered. This is akin to blind compressed sensing \cite{ravishankar2010mr}, where both the image and the sparsifying transform are simultaneously learned during the reconstruction process. Following the work by Xu et al. \cite{xu2020noisy}, we propose training a DL-based denoiser by adding synthetic noise to the image being recovered. The denoiser training process is described in \linref{resides-train} of \alref{resides}. In summary, $\vec{u}_t$ is an intermediate image at iteration $t$, and $\mathcal{I}_i[\vec{u}_t]$ represents the $i^\text{th}$ patch from $\vec{u}_t$. For training purposes, $\mathcal{I}_i[\vec{u}_t]$ and $\mathcal{I}_i[\vec{u}_t] + \mathcal{N}(\vec{0}, s_{t-1}^2 \vec{I})$ act as ``clean'' and ``noisy'' patches, respectively. Here, $\mathcal{N}(\vec{0}, s_{t-1}^2 \vec{I})$ represents complex-valued zero-mean white Gaussian noise with variance $s_{t-1}^2$. The denoiser $\vec{f}(\cdot; \vec{\theta})$, parameterized by $\vec{\theta}$, is trained using $P\geq 1$ patches in a supervised fashion by minimizing the loss $\mathcal{L}(\cdot, \cdot)$, which measures the difference between the denoiser output and clean patches. Once the denoiser is trained, it is then used to denoise the intermediate image $\vec{u}_t$ (\linref{resides-denoise} in \alref{resides}), but this time without the added noise. 
The process of training and applying the denoiser is repeated in each of $T$ iterations or until some convergence criterion is reached.

The strength of the denoiser is controlled by $s_{t}^2$, which plays a role similar to that of regularization strength in CS, i.e., larger $s_{t}^2$ leads to smoother images and smaller $s_{t}^2$ leads to noisier but sharper images. As evident from our preliminary results \cite{liu2022mri}, the value of $s_{t}^2$ also controls the speed of convergence, with larger values preferred in the earlier iterations to speed up convergence and smaller values preferred in later iterations to avoid over-smoothing of the recovered images. To adapt the value of $s_{t}^2$ over iterations, we propose using Morozov's discrepancy principle \cite{wen2011parameter, shastri2020autotuning} (\linref{resides-correction} and \linref{resides-auto} in \alref{resides}). The discrepancy principle is based on the assumption that the squared residual norm, $\|\vec{A}\vec{x}_t - \vec{y}\|_2^2$, monotonically increases with $s_{t}^2$, i.e., more aggressive denoising takes the image away from the least squares solution. By exploiting this monotonic relationship, the discrepancy principle updates $s_{t}^2$ such that the value of $\|\vec{A}\vec{x}_t - \vec{y}\|_2^2$ approaches $M\sigma^2$, which is the expected value of the squared $\ell_2$-norm of $\vec{\eta}$. This is accomplished by using $\frac{M\sigma^2}{\|\vec{A}\vec{x}_t - \vec{y}\|_2^2}$ as a multiplicative correction term, $c_t$. In each iteration, the value of $s_t^2$ is multiplied with $c_t$ to promote the ratio $\frac{M\sigma^2}{\|\vec{A}\vec{x}_t - \vec{y}\|_2^2}$ to be one. The value of $\alpha>0$ (\linref{resides-correction} in \alref{resides}) controls the contributions of the corrective term, with larger values leading to a more rapid adjustment of $\sigma^2_t$. Optionally, a user-defined scalar $\tau > 0$ can be used to provide further control over the final strength of the denoiser, with smaller values generating noisy but sharper images. Note, adjusting $s_t^2$ directly is much more challenging as it needs to be varied over iterations, while the fixed value of $\tau$ can be selected once and then kept constant for a given MRI application. 
%


\begin{algorithm*}[ht!]
\onehalfspacing
\caption{\resides}\label{alg:resides}
\begin{algorithmic}[1]
\Require $\nu>0$, $\tau>0$, $\sigma$, $s_0^2$, $\vec{A}$, $\vec{y}$, $\vec{f}$, $\gamma=\frac{\nu}{\sigma^2}\|\vec{A}\|_2^{2}$, $\vec{x}_0=\vec{A}\herm\vec{y}$, $\vec{z}_0=\vec{Ax}_0-\vec{y}$
\Ensure $\hvec{x}=\vec{x}_T$ \texttt{~~\% final reconstructed image}
\For{$t=1,2,\dots,T$}
        \State{$\vec{u}_t = \vec{x}_{t-1} - \frac{\nu}{\sigma^2}\vec{A}\herm\vec{z}_{t-1}$} \texttt{~~\% intermediate image} 
	\State{$\vec{\theta}_t = \argmin_{\vec{\theta}} \frac{1}{P}\sum\limits_{i=1}^P \mathcal{L}\big(\vec{f}(\mathcal{I}_i[\vec{u}_{t}]+ \mathcal{N}(\vec{0}, s_{t-1}^2 \vec{I}); \vec{\theta}), \mathcal{I}_i[\vec{u}_{t}]\big)$} \texttt{~~\% denoiser training}
    \label{lin:resides-train}
	\State{$\vec{x}_t = \vec{f}(\vec{u}_t; \vec{\theta}_t)$} \texttt{~~\% denoising}
    \label{lin:resides-denoise}
	\State{$\vec{z}_t=\frac{\gamma}{1+\gamma}\vec{z}_{t-1} + \frac{1}{1+\gamma}(\vec{A}(2\vec{x}_t - \vec{x}_{t-1}) - \vec{y})$}
        \State{$c_t=\left(\frac{\tau M \sigma^2}{\|\vec{A}\vec{x}_t-\vec{y}\|^2_2}\right)^\alpha$} \texttt{~~\% correction term}
    \label{lin:resides-correction}
        \State{$s_{t}^2 = c_t s_{t-1}^2$} \texttt{~~\% updating denoiser SNR}
    \label{lin:resides-auto}
\EndFor
\end{algorithmic}
\end{algorithm*}

\subsection{\residem}\label{residem}
There are two major limitations of \resides. First, it requires de novo training of a denoising network in each iteration, which is time-consuming and thus unrealistic for clinical deployment. Second, \resides is a scan-specific method and strictly operates on a single set of undersampled measurements. However, for most MRI applications, more than one set of undersampled measurements is generally available. To reduce computation time at the time of inference and to leverage the availability of multiple sets of undersampled measurements, we propose \residem, which is outlined in \alref{residem}. Here, the tilde annotation on the top of the symbol indicates it is a joint representation of $K \geq 1$ sets of measurements. For example, $\tvec{A}$, $\tvec{y}$, and $\tvec{\sigma}^2$, indicate the forward operator, k-space measurements, and average noise variance, respectively, from $K$ sets of measurements. 

\begin{figure}[ht!]
    \centering
    \includegraphics[width = \columnwidth]{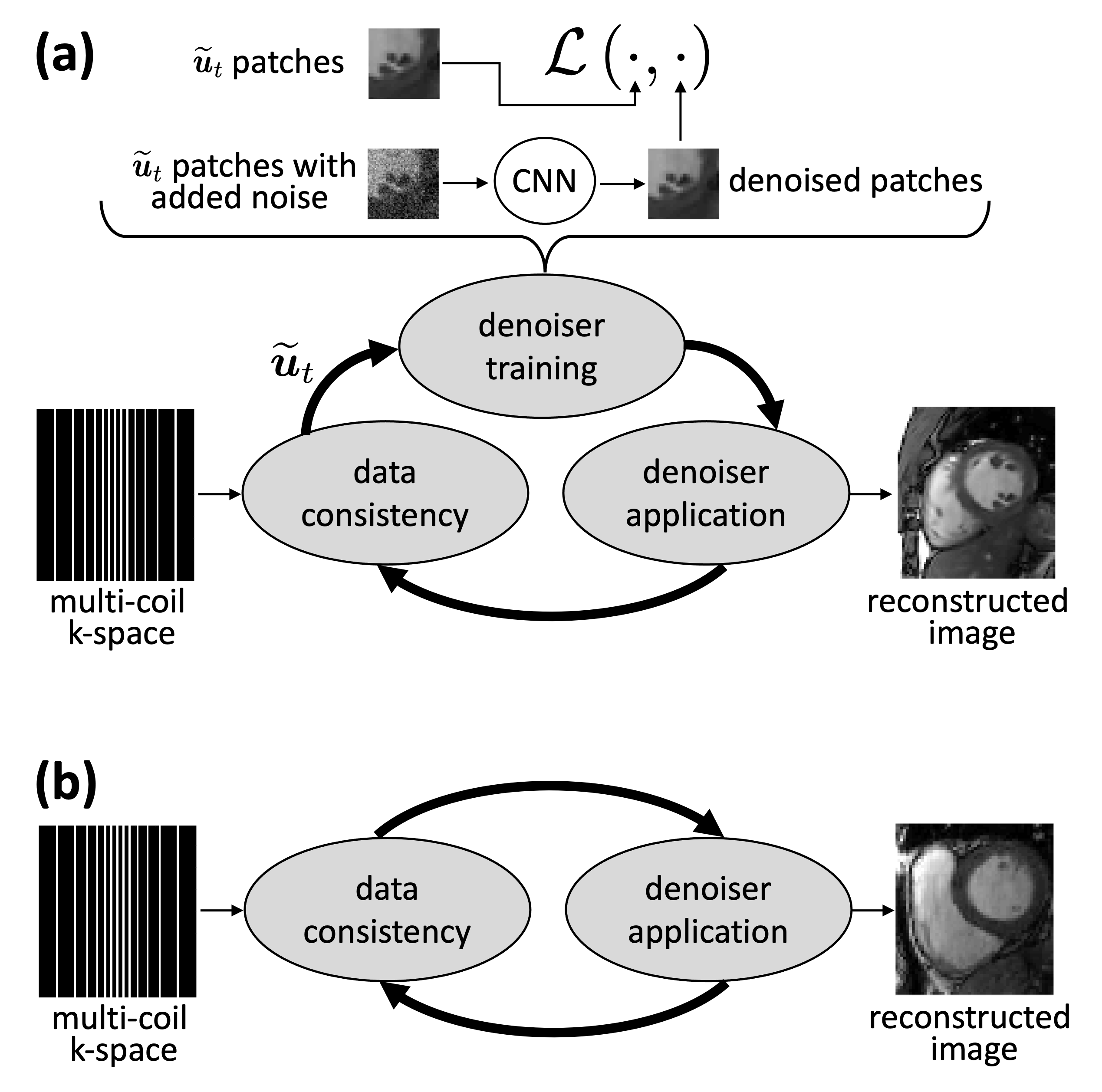}
    \caption{A high-level description of \residem. In the training stage (a), a convolutional neural network (CNN)-based denoiser is trained on patches from intermediate images $\tvec{u}_t$ (\linref{residem-train} in \alref{residem}). The resulting sequence of denoisers is stored. In the inference stage (b), the reconstruction is performed using a PnP method, which conceptually alternates between data consistency and denoising steps. The denoising in (b) is performed by calling the denoisers from (a), without further training.}
    \label{fig:layout}
\end{figure}

\residem is implemented in two stages: training and inference. The training stage is similar to \resides, where both image recovery and denoiser training happen in tandem. However, in contrast to \resides, the denoiser training in \residem is performed using multiple undersampled datasets. For $K=1$, the training phase of \residem is identical to \resides.
Although the training stage of \residem can reconstruct images, its main objective is to store the resulting sequence of denoisers, parameterized by $\{\vec{\theta}_t\}_{t=1}^T$. 
The second stage in \residem is inference, where an unseen undersampled dataset is reconstructed using a PnP algorithm, with the denoising in the $t^{\text{th}}$ iteration of PnP (\linref{residem-apply} in \alref{residem}) performed using the denoiser $\vec{f}(\,\cdot\,;\vec{\theta}_t)$ trained in the first stage. The computational complexity of \residem at the inference stage is similar to that of sparsity-based iterative CS methods. A high-level description of \residem is given in \figref{layout}.

\begin{algorithm*}[ht!]
\onehalfspacing
\caption{\residem}\label{alg:residem}
\begin{algorithmic}[1]
\centerline{\underline{\textbf{Training}}}
\Require $\nu>0$, $\tau>0$, $\tilde{\sigma}^2$, $s_0^2$, $\tvec{A}$, $\tvec{y}$, $\vec{f}$, $\gamma=\frac{\nu}{\tilde{\sigma}^2}\|\tvec{A}\|_2^{2}$, $\tvec{x}_0=\tvec{A}\herm\tvec{y}$, $\tvec{z}_0=\tvec{A}\tvec{x}_0-\tvec{y}$
\Ensure $\{\vec{\theta}_t\}_{t=1}^T$ \texttt{~~\% storing denoisers}
\For{$t=1,2,\dots,T$}
        \State{$\tvec{u}_t = \tvec{{x}}_{t-1} - \frac{\nu}{\sigma^2}\tvec{A}\herm\tvec{z}_{t-1}$} \texttt{~~\% intermediate image}
	\State{$\vec{\theta}_t = \argmin_{\vec{\theta}} \frac{1}{P}\sum\limits_{i=1}^P \mathcal{L}(\vec{f}(\mathcal{I}_i[\tvec{u}_{t}]+ \mathcal{N}(\vec{0}, s_{t-1}^2 \vec{I}); \vec{\theta}), \mathcal{I}_i[\tvec{u}_{t}])$} \texttt{~~\% denoiser training}
    \label{lin:residem-train}
	\State{$\tvec{x}_t = \vec{f}(\tvec{{u}}_t; \vec{\theta}_t)$} \texttt{~~\% denoising}
	\State{$\tvec{z}_t=\frac{\gamma}{1+\gamma}\tvec{{z}}_{t-1} + \frac{1}{1+\gamma}(\tvec{A}(2\tvec{x}_t - \tvec{x}_{t-1}) - \tvec{y})$}
        \State{$c_t=\left(\frac{\tau M \tilde{\sigma}^2}{\|\tvec{A}\tvec{x}_t-\tvec{y}\|^2_2}\right)^\alpha$} \texttt{~~\% correction term}
        \label{lin:residem-correction}
	\State{$s_{t}^2 = c_t s_{t-1}^2$} \texttt{~~\% updating denoiser SNR}
        \label{lin:residem-auto}
\EndFor

\centerline{\underline{\textbf{Inference}}}
\Require $\nu>0$, $\sigma$, $\vec{A}$, $\vec{y}$, $\vec{f}$, $\gamma=\frac{\nu}{\sigma^2}\|\vec{A}\|_2^{2}$, $\vec{x}_0=\vec{A}\herm\vec{y}$, $\vec{z}_0=\vec{Ax}_0-\vec{y}$, $\{\vec{\theta}_t\}_{t=1}^T$
\Ensure $\hvec{x}\gets\vec{x}_t$ \texttt{~~\% final reconstructed image}
\For{$t=1,2,\dots,T$}
        \State{$\vec{u}_t = \vec{x}_{t-1} - \frac{\nu}{\sigma^2}\vec{A}\herm\vec{z}_{t-1}$} \texttt{~~\% intermediate image}
	\State{$\vec{x}_t = \vec{f}(\vec{u}_t; \vec{\theta}_t)$} \texttt{~~\% denoising}
    \label{lin:residem-apply}
	\State{$\vec{z}_t=\frac{\gamma}{1+\gamma}\vec{z}_{t-1} + \frac{1}{1+\gamma}(\vec{A}(2\vec{x}_t - \vec{x}_{t-1}) - \vec{y})$}
\EndFor
\end{algorithmic}
\end{algorithm*}

\subsection{Study I--Brain MRI}
In this study, \resides and \residem were evaluated on T1- and T2-weighted images from the fastMRI dataset \cite{zbontar2018fastmri}. For each contrast, twenty-two sets of multi-coil measurements were used. 
All T1 images were cropped to $320\times 320$, and all T2 images were cropped to $384\times 304$. \textr{Since the noise pre-scan was not available for the fastMRI data, noise pre-whitening was not applied.} The multi-coil k-space data were compressed to eight virtual coils \cite{buehrer2007array}. The data were retrospectively downsampled at $R=4$ using two realistic Cartesian sampling masks, i.e., a 1D pseudo-random mask with a 32-line wide ACS region (M1) and a 1D random mask with a 32-line wide ACS region (M2). 
The M1 mask was kept fixed across training and testing sets of measurements, while a different M2 mask was randomly drawn for each training and testing instance. The coil sensitivity maps were estimated using ESPIRiT \cite{uecker2014espirit}. Sixteen out of the 22 sets of measurements were used for network training in \residem and SSDU. 
Five measurement sets were used for performance evaluation and for comparing \resides and \residem with CS, PnP-BM3D, ConvDecoder \cite{darestani2021accelerated}, and SSDU \cite{yaman2020self}. The CS reconstruction employed $\ell_1$ regularization in the wavelet domain and was implemented using SigPy \cite{Ong2023}. 
The PnP-BM3D reconstruction was implemented using \alref{pds}, with $\vec{f}(\cdot)$ in \linref{pds-denoise} of \alref{pds} representing a call to the BM3D denoiser \cite{danielyan2011bm3d}. 
\textr{To benchmark against a well-established supervised learning method, we further compare \residem with a VarNet model pre-trained on brain fastMRI data using rate-4 equispaced (ES) sampling and 8\% ACS, resulting in a net acceleration rate of 3.2 \cite{sriram2020end}.}


\textr{For each method, one set of measurements was used for manual parameter tuning to maximize peak SNR (PSNR).} These parameters included: regularization strength, $\lambda$, for CS, denoising level for BM3D in PnP-BM3D, number of iterations and input size for ConvDecoder, cardinality of the loss mask and number of epochs for SSDU, and parameters $\alpha$ and $\tau$, number and the size of patches, and total number of iterations, $T$,  for \resides and \residem. For CS, the reconstruction process was repeated with the regularization set at $\lambda /2$ and $2\lambda$ to assess the impact of regularization strength on the image quality.

\subsection{Study II--MRXCAT Perfusion Phantom}
Twenty-two sets of perfusion image series from the MRXCAT digital phantom were considered in this study \cite{wissmann2014mrxcat}. Each set of measurements was cropped to $112 \times 168$ pixels, with $32$ frames and $4$ receive coils. All data were retrospectively downsampled at $R=4$ using a 1D pseudo-random Cartesian sampling mask (M3) \cite{joshi2022technical}. Due to the interleaving nature of M3, the ACS region was not acquired for individual frames, and the fully sampled time-averaged k-space was used to estimate coil sensitivity maps. To simulate realistic data, circularly symmetric zero-mean white Gaussian noise was added to k-space measurements to generate a signal-to-noise ratio of approximately 12 dB. Sixteen sets of measurements were used to train \residem, and five sets of measurements were used for performance evaluation and for comparing \resides and \residem with CS and PnP-BM4D. The CS reconstruction employed $\ell_1$ regularization in the spatiotemporal wavelet domain \cite{chen2019sparsity}. The PnP-BM4D reconstruction was implemented using \alref{pds}, with $\vec{f}(\cdot)$ in \linref{pds-denoise} of \alref{pds} representing a call to the BM4D denoiser \cite{xu2019new}. As described in the previous study, one set of measurements was used to manually optimize free parameters in each method.



\subsection{Study III--First-Pass Perfusion}
This study included 22 first-pass perfusion image series from patients clinically referred for a cardiac MRI exam at our institute. All measurements were performed on a commercial 1.5T scanner (MAGNETOM Sola, Siemens Healthcare, Erlangen, Germany) with a fast low angle shot (FLASH) sequence using echo-planar imaging (EPI) readout. The data were collected in three different views, i.e., short-axis, two-chamber, and four-chamber views. The other imaging parameters were: flip angle 25 degrees, temporal footprint $75.48$ to $99.36$ ms, matrix size $144\times 108$ to $144\times 144$, field of view $360 \times 270$ to $420 \times 380$, echo train length $4$, echo spacing $6.06$ to $6.29$ ms, slice thickness 8 to 10 mm, and a number of frames 60. The images were prospectively undersampled in the $k_x$-$k_y$ domain with an acceleration rate of two and uniform undersampling that was interleaved across time. \textr{Noise pre-whitening was applied by using the noise pre-scan, and the multi-coil k-space data were compressed to eight virtual coils \cite{buehrer2007array}.} Sixteen sets of measurements were used to train \residem, and five sets of measurements were used for performance evaluation and comparison with PnP-BM4D and CS with $\ell_1$ regularization in the spatiotemporal wavelet domain \cite{chen2019sparsity}. Again, one set of measurements was used to manually optimize free parameters. 


\subsection{Quality Assessment}
In Studies I and II, where the fully sampled reference was available, image quality was assessed using the structural similarity index (SSIM) and \textr{PSNR in dB, defined as $20\log_{10}\left(\sqrt{N}|\vec{x}|_\text{max} / \|\vec{x} - \hvec{x}\|_2\right)$, with $|\vec{x}|_{\text{max}}$ being the maximum absolute value in $\vec{x}$.} For Study III, where the fully sampled reference was not available, each perfusion image series was blindly scored by three expert reviewers, including two cardiologists, each with more than ten years of experience in cardiac MRI. Each image series, presented as a movie, was scored on a five-point Likert scale (1: non-diagnostic, 2: poor, 3: adequate, 4: good, 5: excellent) in terms of overall image quality.

\subsection{Implementation of \reside}
\begin{figure}[ht!]
    \begin{center}
    \includegraphics[width = \columnwidth]{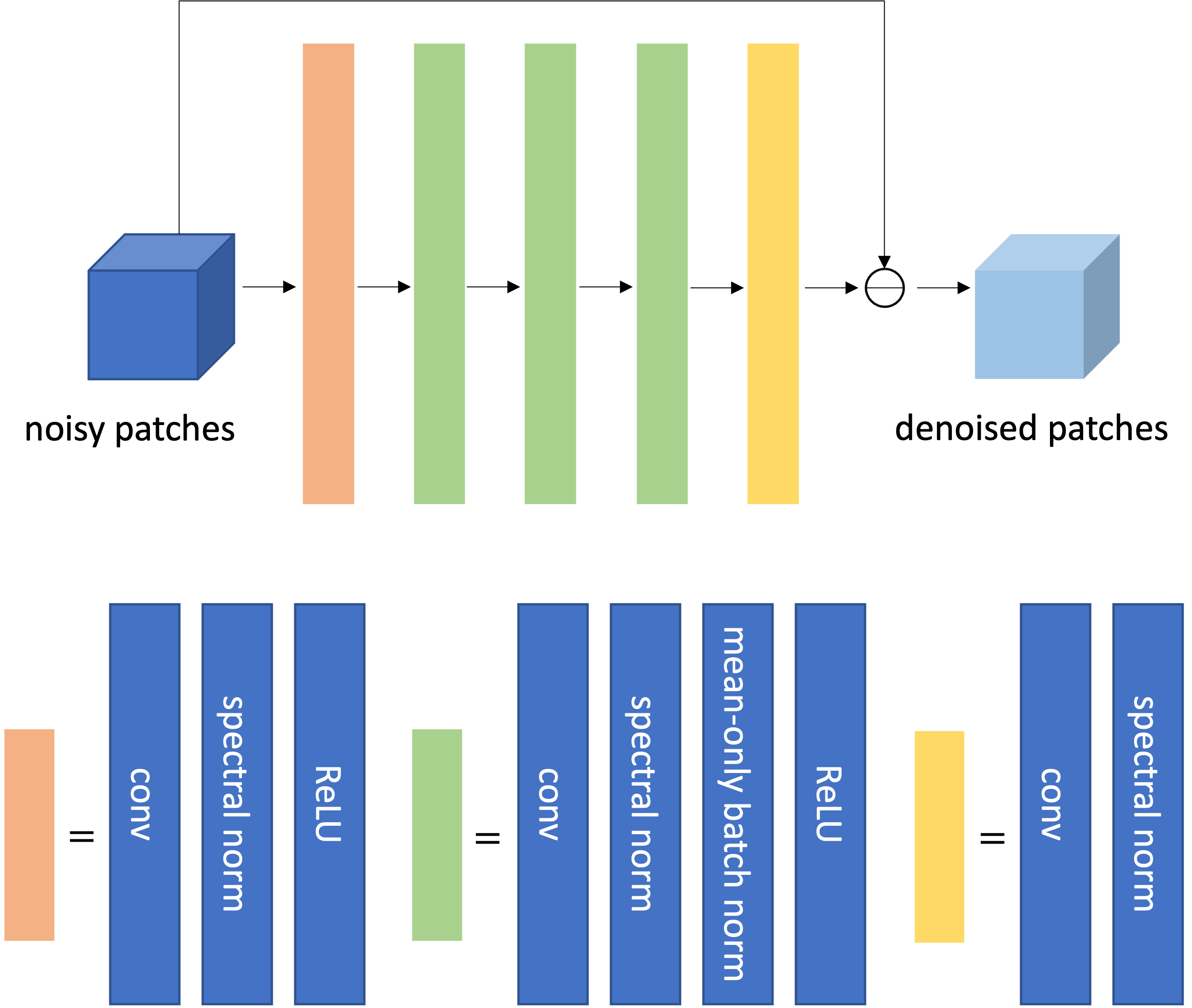}
    \end{center}
    \caption{The denoiser architecture used in \resides and \residem. Here, ``conv'' represents 2D convolutional layer for Study I and 3D convolutional layer for Studies II and III.}
    \label{fig:network}
\end{figure}

In Study I, randomly positioned $P=576$ patches and $P=2,\!306$ patches were used to train the denoiser in \resides and \residem, respectively. For \residem, the $2,\!306$ patches were evenly distributed across the 16 training images. The patch size was fixed at $64 \times 64$. For Studies II and III, randomly positioned $P=288$ patches and $P=4,\!608$ patches were used to train the denoiser in \resides and \residem, respectively. For \residem, the $4,\!608$ patches were again evenly distributed across the 16 training image series. The patch size was fixed at $64 \times 64 \times 20$. In all three studies, the locations of patches were randomly shuffled from one epoch to the next. The mean squared error was used as a cost function to train the denoiser. The real and imaginary parts were split into two channels. The architecture of the denoiser is shown in \figref{network}. Each convolutional layer had 128 kernels of size $3 \times 3$ for Study I and size $3 \times 3 \times 3$ for Studies II and III. We used Adam optimizer with the learning rate $10^{-3}$ for Study I and $10^{-4}$ for Studies II and III. The training in \residem was performed on an NVIDIA RTX 2080 Ti for Study I and an NVIDIA RTX 3090 for Studies II and III. For Study I, the measurement noise variance $\sigma^2$ was approximated from the outer fringes of k-space. For Study II, the noise was synthetically added; therefore, the value of $\sigma^2$ was precisely known. \textr{For Study III, the value of $\sigma^2$ was inferred from the data scaling factor applied after noise pre-whitening.} The value of $\tau$ was set at 0.65, 0.9, and 1.15 for Studies I, II, and III, respectively, and the value of $\alpha$ was set at 0.1 for all studies. The value of $s_0^2$ was set such that the initial SNR for the denoiser training was 5 dB. The number of iterations, $T$, for \resides and \residem was set at 80 for all three studies. Within each iteration, the denoiser was trained for a total of ten epochs. The code for \resides and \residem is available here \href{https://github.com/OSU-MR/reside}{https://github.com/OSU-MR/reside}

\section{Results}\label{sec:res}
\subsection{Study I--Brain MRI}
\begin{figure}[ht!]
    \begin{center}
    \setlength{\unitlength}{1cm} 
    \begin{picture}(0,0)
        \put(-0.2,2.6){\rotatebox{90}{\textcolor{blue}{$c_t$}}} 
        \put(7,3.2){\rotatebox{-90}{\textcolor{red}{PSNR}}} 
        \put(2.8,-0.3){\sf{iteration,} $t$} 
    \end{picture}
    \includegraphics[width = 0.9\columnwidth]{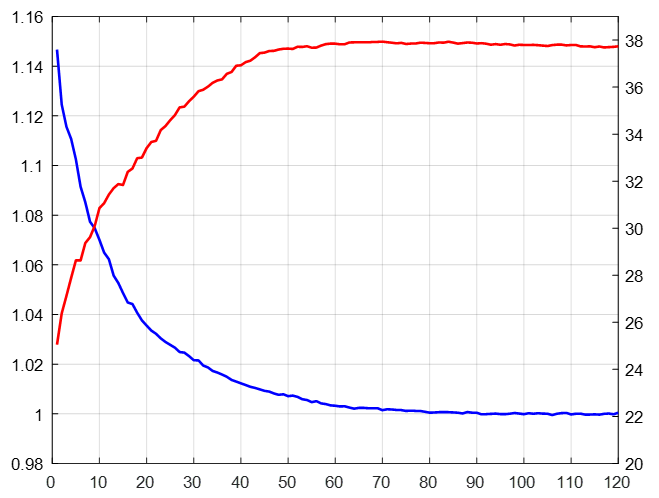}
    \vspace{0.5cm} 
    \end{center}
    \caption{Peak SNR (PSNR) and the multiplicative correction term, $c_t$, as a function of iteration number, $t$. These representative curves originate from one of the T1-weighted images reconstructed using \resides.}
    \label{fig:conv}
\end{figure}
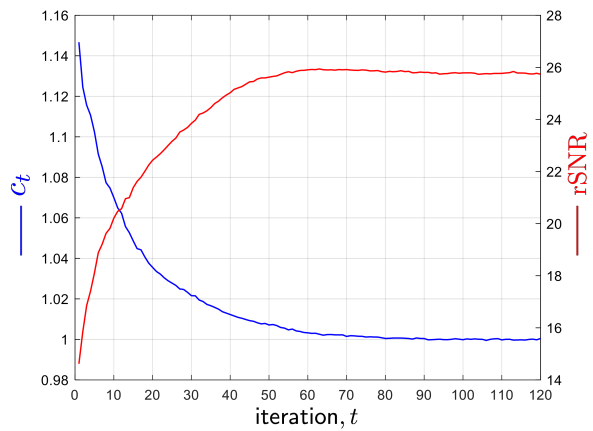

\begin{figure*}[ht!]
    \begin{center}
    \includegraphics[width = \textwidth]{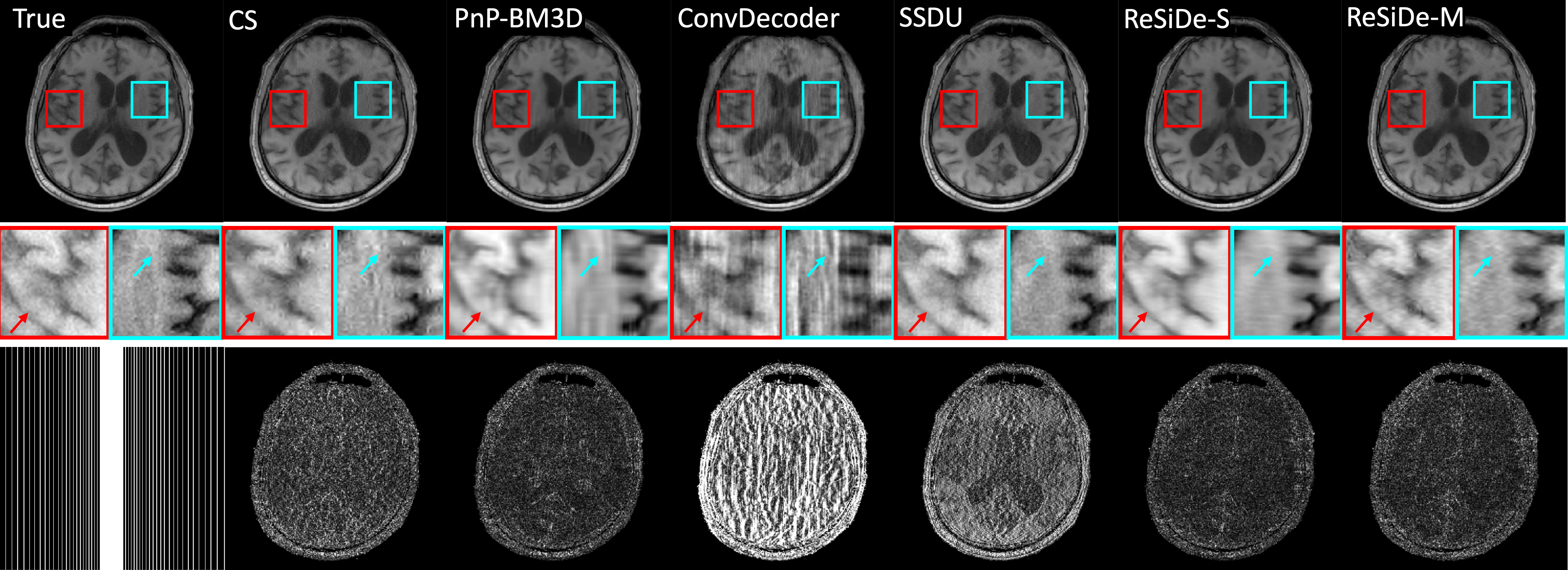}
    \end{center}
    \caption{An example showing reconstruction of a T1-weighted image with sampling mask M1. To highlight differences, the second row magnifies two areas in the brain. The arrows point to features where some of the methods show artifacts or blurring. The third row shows the sampling mask (left) and the absolute error maps after five-fold amplification.}
    \label{fig:t1_m1}
\end{figure*}

\begin{figure*}[ht!]
    \begin{center}
    \includegraphics[width = \textwidth]{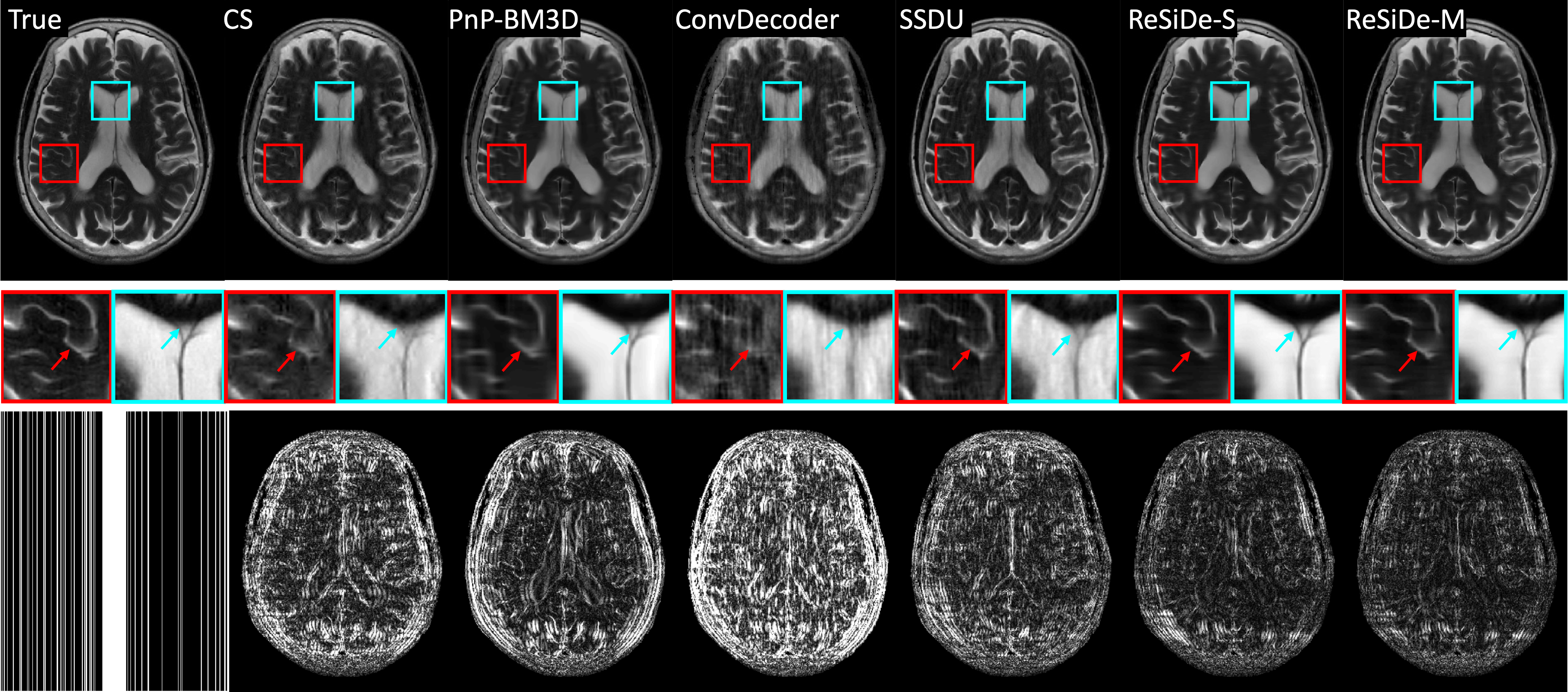}
    \end{center}
    \caption{An example showing reconstruction of a T2-weighted image with sampling mask M2. To highlight differences, the second row magnifies two areas in the brain. The arrows point to features where some of the methods show artifacts or blurring. The third row shows the sampling mask (left) and the absolute error maps after five-fold amplification.}
    \label{fig:t2_m2}
\end{figure*}

\figref{conv} shows a representative example where the values of PSNR and the multiplicative correction term, $c_t$, are plotted as a function of the number of iterations for \resides.
\figref{t1_m1} and \figref{t2_m2} display examples of reconstructed T1- and T2-weighted images using undersampling masks M1 and M2, respectively. The first row presents the true image obtained from fully sampled k-space, alongside the images reconstructed by CS, PnP-BM3D, ConvDecoder, SSDU, VarNet, \resides, and \residem. The second row features two magnified regions from the images in the first row. The arrows highlight areas where visible artifacts or blurring is present in some of the reconstructed images. In the third row, the leftmost panel shows the undersampling masks, while the remaining panels depict error maps from various reconstructions after a five-fold amplification. The top section of \tabref{brain} summarizes PSNR and SSIM values averaged over five T1- and T2-weighted images employing M1 and M2 masks. The PSNR/SSIM of CS reduced from $32.85/0.921$  to $32.02/0.907$ and $32.40/0.917$ when the value of the regularization strength was changed to $\lambda /2$ and $2\lambda$, respectively, indicating that the value of $\lambda$ selected in CS was optimal or near-optimal.

\begin{figure}[ht!]
    \begin{center}
    \includegraphics[width = \columnwidth]{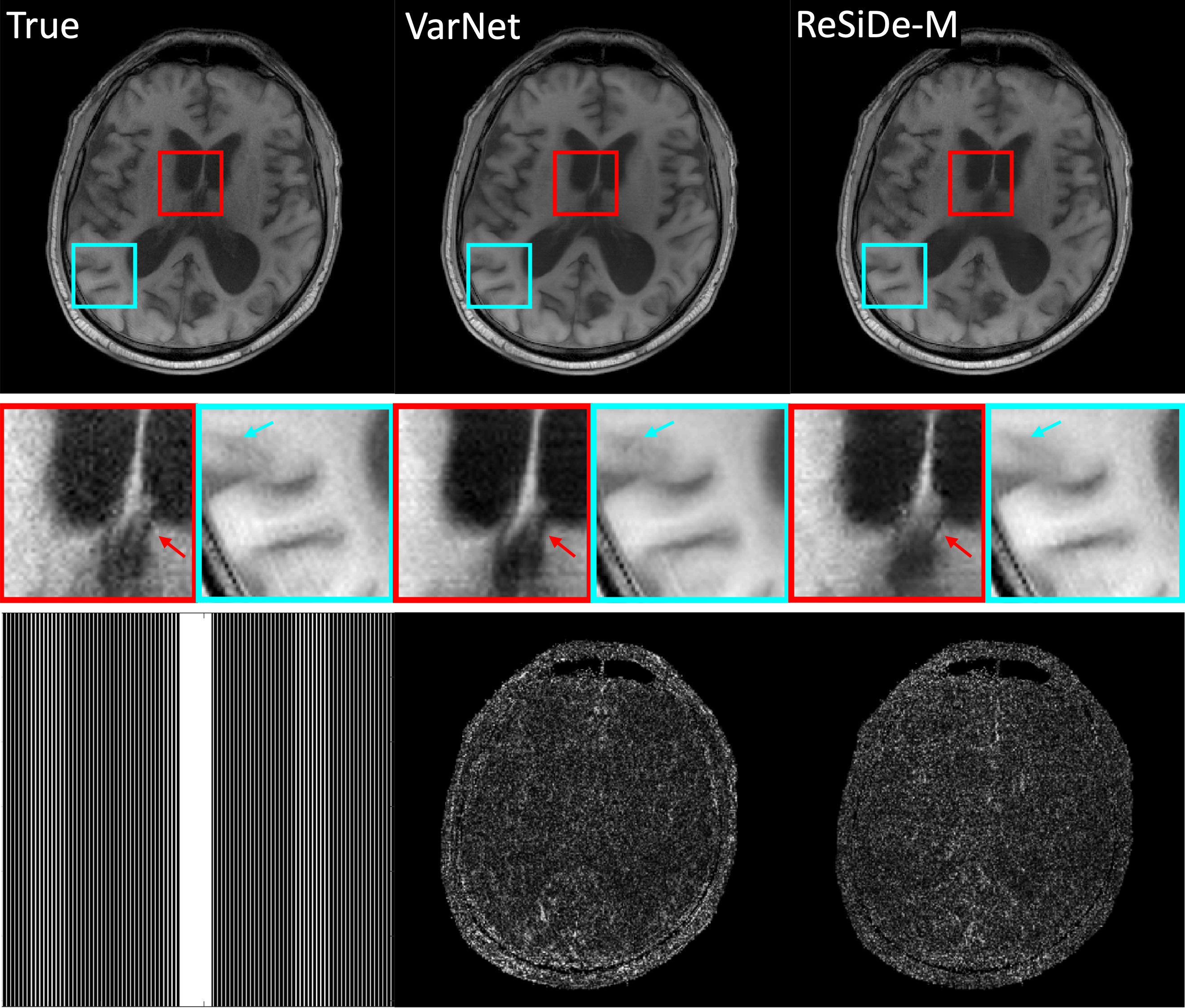}
    \end{center}
    \caption{\textr{An example showing reconstruction of a T1-weighted image with equispaced (ES) sampling. To highlight differences, the second row magnifies two areas in the brain. The arrows point to features where the two reconstructions show subtle differences. The third row shows the sampling mask (left) and the absolute error maps after five-fold amplification.}}
    \label{fig:t1_uni}
\end{figure}

\textr{The comparison between \residem and VarNet is summarized in \tabref{varnet}. Overall, \residem compares favorably to VarNet. Despite being trained on the entire fastMRI dataset, the PSNR of VarNet, when averaged over ten T1 and T2 test images, is only 0.27 dB higher than that of \residem. \figref{t1_uni} shows an example of reconstructed T1-weighted image. Although VarNet reconstruction appears slightly clearer, the differences between the reconstructions are subtle. It is worth mentioning that when the VarNet trained on ES sampling was applied to M1 and M2 sampling patterns, its resulting PSNR value of 32.97 dB was 3.78 dB worse than the value of 36.75 dB achieved by \residem. This is because the performance of end-to-end methods, like VarNet, typically degrades when the forward model at the inference stage is different from the one used during training. In contrast, self-supervised methods like ours circumvent this limitation. }

\setlength{\arrayrulewidth}{1.2pt}
\begin{table*}[ht!]
\centering
\resizebox{\textwidth}{!}{%
\renewcommand{\arraystretch}{1.5}
\begin{tabular}{!{\color{black}\vrule}c|c|c|c|c|c|c|c|c!{\color{black}\vrule}}
\hline
\rowcolor{gray!50}
\textbf{Image} & \textbf{Samp} & \textbf{CS} & \textbf{PnP-BMXD} & \textbf{ConvDecoder} & \textbf{SSDU} & \textbf{ReSiDe-S} & \textbf{ReSiDe-M} \\ \hline\hline
Brain T1 & M1 & 34.55/0.932 & 35.84/0.949 & 26.15/0.837 & 33.28/0.946 & 35.89/0.947 & $\bm{36.26/0.950}$ \\ \hline
Brain T1 & M2 & 30.07/0.874 & 30.23/0.923 & 25.63/0.782 & 31.34/0.925 & $\bm{34.47/0.947}$ & 34.32/0.946 \\ \hline
Brain T2 & M1 & 34.64/0.944 & 36.54/0.960 & 27.01/0.841 & 32.99/0.943 & 39.06/0.972 & $\bm{39.09/0.973}$ \\ \hline
Brain T2 & M2 & 32.13/0.932 & 33.47/0.940 & 26.73/0.824 & 31.35/0.934 & $\bm{37.37/0.976}$ & 37.34/0.975 \\ \hline
\rowcolor{gray!25}
Brain Avg & M1/M2 & 32.85/0.921 & 34.02/0.943 & 26.38/0.821 & 32.24/0.937 & 36.70/0.961 & $\bm{36.75/0.961}$ \\ \hline\hline
MRXCAT & M3 & 34.91/0.877 & 36.67/0.900 & -- & -- & 38.79/0.918 & $\bm{38.98/0.921}$ \\ \hline
\end{tabular}%
}
\caption{Image quality metrics for Study I (top) and Study II (bottom). In each cell, the first number represents PSNR in dB and the second number represents SSIM, both averaged over five test samples. The best value in each row is highlighted in bold font. The ``Brain Avg'' row represents the average of the preceding four rows. BMXD represents BM3D for Study I and BM4D for Study II.}
\label{tab:brain}
\end{table*}

\subsection{Study II--MRXCAT Perfusion Phantom}
\begin{figure*}[ht!]
    \begin{center}
    \includegraphics[width = \textwidth]{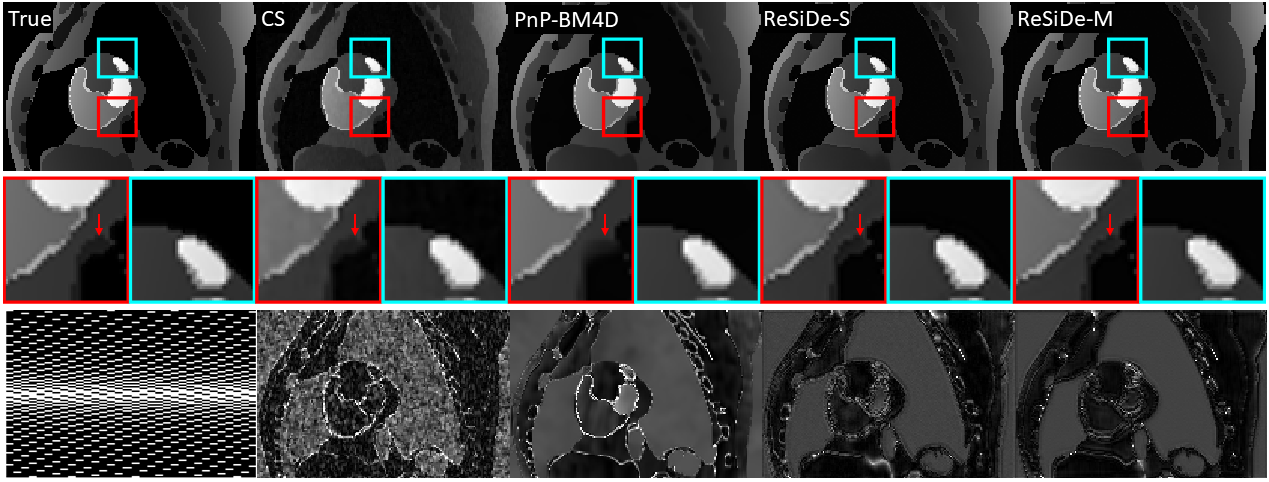}
    \end{center}
    \caption{A representative frame from MRXCAT perfusion phantom reconstruction. The second row magnifies two areas of the phantom. The arrows point to features where some of the methods show blurring. The third row shows the sampling mask (left) in the phase-encoding (vertical) and temporal (horizontal) dimensions and the absolute error maps after five-fold amplification.}
    \label{fig:mrxcat_01}
\end{figure*}

\figref{mrxcat_01} presents a representative frame from reconstructions of an MRXCAT perfusion phantom. The first row displays the true image derived from fully sampled k-space, as well as images reconstructed using CS, PnP-BM4D, \resides, and \residem methods. The middle row shows a magnification of two selected regions. The arrows emphasize details that are partially or entirely lost in some of the reconstructed images. In the bottom row, the leftmost panel shows the undersampling masks in phase encoding (vertical) and temporal (horizontal) dimensions. The readout dimension, which is not shown, is fully sampled. The remaining panels in the third row depict error maps after a five-fold amplification. The last row in \tabref{brain} summarizes PSNR and SSIM values averaged over five MRXCAT datasets with M3 mask for CS, PnP-BM4D, \resides, and \residem.

\begin{table}[ht!]
\centering
\renewcommand{\arraystretch}{1.5}
\textr{
\begin{tabular}{!{\color{black}\vrule}c|c|c|c!{\color{black}\vrule}}
\hline
\rowcolor{gray!50}
\textbf{Image} & \textbf{Samp} & \textbf{VarNet} & \textbf{ReSiDe-M} \\ \hline\hline
Brain T1 & ES &  36.02/0.961 & $\bm{37.28/0.962}$ \\ \hline
Brain T2 & ES &  $\bm{40.98/0.985}$ & 39.17/0.973 \\ \hline
\rowcolor{gray!25}
Brain Avg & ES  & $\bm{38.50/0.973}$ & 38.23/0.968 \\ \hline
\end{tabular}}%
\caption{\textr{Comparison of \residem and VarNet for equispaced (ES) sampling.} }
\label{tab:varnet}
\end{table}

\begin{table}[ht!]
\scriptsize
\centering
\renewcommand{\arraystretch}{1.5}
\begin{tabular}{|C{0.9cm}|C{0.9cm}|C{1.1cm}|C{1.2cm}|C{1.2cm}|}
\hline
\rowcolor{gray!50}
   & \textbf{CS} & \textbf{PnP-BM4D} & \textbf{\resides} & \textbf{\residem} \\ \hline \hline
E1  & $3.2$ & $4.0$ & $4.4$ & $\mathbf{5.0}$ \\ \hline
E2  & $2.0$ & $3.0$ & $4.0$ & $\mathbf{4.6}$ \\ \hline
E3  & $2.0$ & $3.4$ & $3.6$ & $\mathbf{3.6}$ \\ \hline
Avg  & $2.4$ & $3.5$ & $4.0$ & $\mathbf{4.4}$ \\ \hline
\end{tabular}
\caption{Image quality scores from three expert reviewers (E1, E2, and E3) on a five-point Likert scale (5: best, 1: worst), averaged over five perfusion image series.}
\label{tab:perfusion}
\end{table}

\subsection{Study III--First-Pass Perfusion}
\begin{figure*}[ht!]
    \begin{center}
    \includegraphics[width = \textwidth]{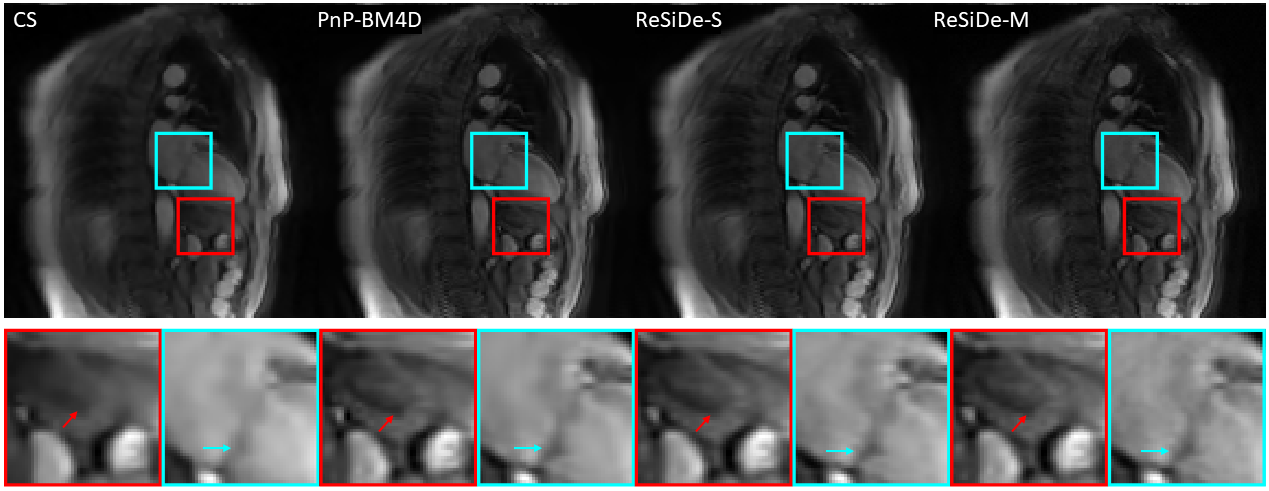}
    \end{center}
    \caption{A representative two-chamber frame from one of the first-pass perfusion image series. The first row shows the entire frame, while the second shows two magnified areas from the frame. The visible loss of detail in CS is highlighted with arrows.}
    \label{fig:perfusion17}
\end{figure*}

\begin{figure*}[h!]
    \begin{center}
    \includegraphics[width = \textwidth]{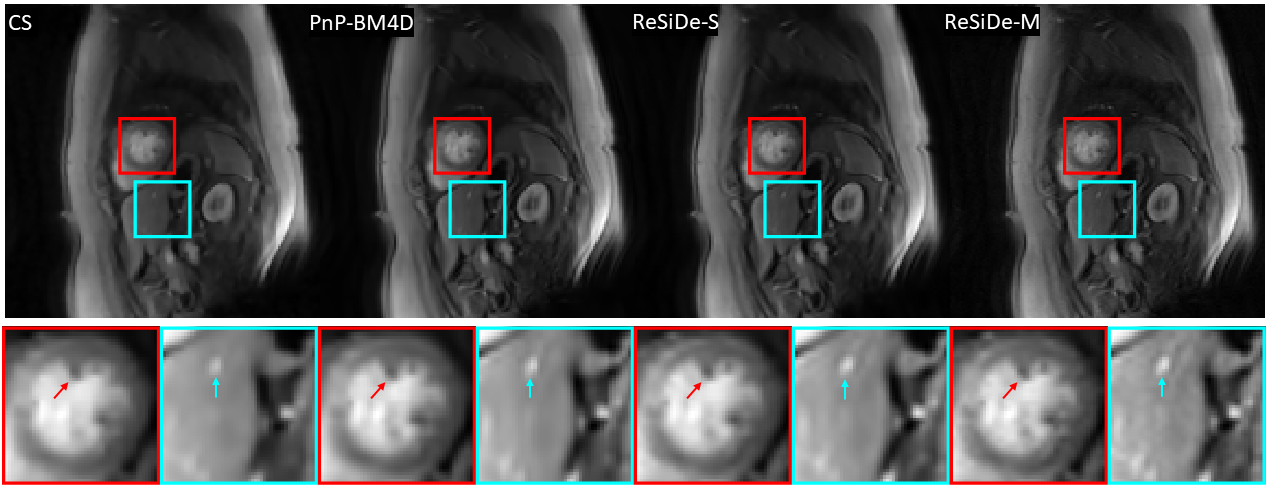}
    \end{center}
    \caption{A representative short-axis frame from another first-pass perfusion image series. The first row shows the entire frame, while the second shows two magnified areas from the frame. The visible loss of detail in CS is highlighted with arrows.}
    \label{fig:perfusion19}
\end{figure*}

\figref{perfusion17} and \figref{perfusion19} show representative frames from two different first-pass perfusion image series. Reconstructions from CS, PnP-BM4D, \resides, and \residem are shown. The top row shows the entire frame, and the bottom row displays two magnified regions from the images in the first row. The arrows highlight the details that are partially lost in some of the reconstructed images. In \figref{perfusion17}, the cyan arrows point to the leaflets of the mitral valve. In \figref{perfusion19}, the red arrows point to the papillary muscles in the left ventricle. For CS, PnP-BM4D, \resides, and \residem, \tabref{perfusion} provide the image quality scores averaged over five image series from three cardiac MRI experts, including two cardiologists.

\section{Discussion}

In this work, we present an SSDL method, called \reside, for MRI reconstruction. Like PnP methods, \reside integrates a denoiser into the reconstruction process. However, PnP methods use pre-trained denoisers, while \reside iteratively trains the denoiser on the images being recovered. We present two variations of \reside, i.e., \resides and \residem. \resides is scan-specific and utilizes only a single set of undersampled measurements. The necessity to train the network in each iteration makes \resides computationally slow. 
\textr{In contrast, \residem operates on multiple sets of undersampled measurements and thus employs a more comprehensive training of the denoiser using patches from multiple images.} The denoisers trained in \residem are stored and then utilized in a PnP algorithm without further training. The computation burden of \residem after the training stage is comparable to CS-based iterative methods.

Validation of \resides and \residem is carried out in three datasets, i.e., T1- and T2-weighted images from fastMRI (Study I), digital perfusion image series from MRXCAT (Study II), and first-pass perfusion data collected from patients (Study III). In Study I, we compare \resides and \residem with other methods that do not require fully sampled data, including CS, PnP-BM3D, ConvDecoder, and SSDU. As summarized in \tabref{brain}, \residem and \resides outperform competing methods in terms of PSNR and SSIM, with more than 2.5 dB improvement over PnP-BM3D and SSDU. 
All methods perform better with the M1 mask compared to the M2 mask. This is not surprising because M1 is pseudo-random and thus avoids large sampling gaps in k-space. Also, M1 is kept consistent between training and testing, while a different M2 mask is used for each training and testing instance. Nonetheless, the smaller performance gap between M1 and M2  for \reside highlights its ability to generalize when the sampling patterns between training and testing are different. 
Two examples of reconstructed images are shown in \figref{t1_m1} and \figref{t2_m2}. Compared to PnP-BM3D, ConvDecoder, and SSDU, both \resides and \residem exhibit fewer artifacts while preserving fine details. The artifacts are particularly pronounced in ConvDecoder. Despite our best efforts to optimize the code provided by the original authors here \cite{Darestani2022}, we were unable to improve the performance of ConvDecoder. The performance of CS was inferior to \resides and \residem by more than 3 dB. To explore if this performance gap is due to the suboptimal selection of the regularization strength in CS, we repeated the CS reconstruction at lower ($\lambda/2$) and higher ($2\lambda$) regularization strengths. However, the performance of CS degraded for both those choices, indicating that the inferior performance of CS cannot be attributed to suboptimal selection of $\lambda$.

The PSNR and SSIM numbers from Study II followed a trend similar to that of Study I, with both \resides and \residem outperforming other methods. \figref{mrxcat_01} shows a representative frame. As shown in the error map, \resides and \residem preserve edge information more effectively, whereas CS and PnP-BM4D show structured errors around the edges. The performance gap is more than 1.5 dB between \reside and PnP-BM4D and more than 3 dB between \reside and CS. 

In Study III, where the image series were subjectively evaluated by three expert readers, \residem consistently outperformed other methods, with \resides being the second best. The example images provided in \figref{perfusion17} and \figref{perfusion19} illustrate that the \reside methods are more effective in preserving small details, e.g., mitral valve leaflets or papillary muscles, especially compared to CS.

The performance difference between \resides and \residem, in terms of PSNR and SSIM, was marginal in Studies I and II. However, we observed that \residem images have more texture and appear sharper than those from \resides. We conjecture that this behavior is related to more expansive training of the denoiser in \residem. In addition to the superior image quality, \residem also offers a computation advantage over \resides and many other SSDL methods. For example, in Study I, \resides and \residem trainings took 36 minutes and 140 minutes, respectively. However, at the inference stage, the reconstruction from \residem took only 11 seconds per image. In comparison, the training and inference from SSDU took 127 minutes and 3.3 seconds, respectively. The inference time for \residem for Studies II and III was 25 and 37 seconds, respectively, per image series. In comparison, PnP-BM4D took 40 minutes and 110 minutes for each image series in Studies II and III, respectively.

The performance of \reside depends on $s_t^2$, which controls the training SNR of the denoiser. We employ the discrepancy principle to adapt the value of $s_t^2$. \figref{conv} shows representative curves for PSNR and $c_t$ for one the T1-weighted images. As expected, the value of the corrective term, $c_t$, converges to one after approximately 80 iterations, which implies that $s_t^2$ also converges to a fixed value. Likewise, the PSNR reaches its maximum value at around the 60 iteration mark, with marginal or no drop-off in PSNR for iterations beyond 60. Similar trends were observed for other datasets in Studies I, II, and III. The value of $\alpha$ had a marginal impact on the final PSNR but it did affect the rate of convergence. In particular, larger values of $\alpha$ led to faster convergence but with PSNR and $c_t$ curves showing more oscillations, especially in the earlier iterations. Therefore, the value of $\alpha$ was conservatively set at 0.1. \textb{Also, the performance of \reside does not vary significantly with the number of patches as long as that number is large enough. The values of $\tau$ and patch size do affect image quality and were optimized on a separate set of measurements in each study.}

This study has several limitations. First, our current implementation requires saving a denoiser in each iteration of the training process. Saving a large number of denoisers can be memory intensive, especially if larger networks are employed. Future work could explore saving the denoisers less frequently, e.g., after every tenth iteration. Second, we have used a denoiser architecture that is based on the residual learning approach proposed in 2017 \cite{zhang2017beyond}. \textb{It is possible that more recent network architectures that utilize attention mechanisms \cite{niu2021review} can further improve the performance of \reside.} \textr{Third, the denoiser training used in \reside does not explicitly use the g-factor information, and using this information can further improve the performance \cite{xue2024imaging}.} Fourth, although using the discrepancy principle eliminates the need to precisely schedule the denoiser strength \cite{liu2022mri}, both \resides and \residem still require selecting an appropriate value for $\tau$. For the studies presented, we manually selected one value of $\tau$ for each application. It is not clear whether this value of $\tau$ will remain reasonable if imaging parameters, e.g., spatial resolution or measurement SNR, change significantly within each application. However, \reside shares this limitation with CS and PnP methods, which also require selecting the regularization or denoising strength. Fourth, we observed \resides and \residem to converge to a fixed point in all three represented studies. However, a more formal convergence analysis for \reside is currently missing and beyond the scope of this proof-of-concept work.

\section{Conclusions}
We have presented two self-supervised methods for MRI reconstruction: \resides and \residem. \resides offers a scan-specific implementation where a single set of undersampled measurements is used for denoiser training and image recovery. In contrast, \residem trains a denoiser from multiple undersampled measurements and utilizes that denoiser during inference without further training. Our validation studies, which used data from brain MRI, perfusion phantom, and first-pass perfusion, demonstrate that \resides and \residem outperform other self-supervised or unsupervised methods in terms of both qualitative and quantitative metrics. Compared to \resides, \residem also offers slightly better image quality and much faster inference.

\backmatter




\section*{Statements and Declarations}
\textbf{Funding} This work was funded by NIH grants R01-EB029957, R01-HL151697, and R01-HL135489.

\noindent
\textbf{Ethics approval} For the human subject data, approval was granted by the Institutional Review Board (IRB) at The Ohio State University (2019H0076).

\noindent
\textbf{Consent to participate} Informed consent was obtained from all individual participants included in this work.

\noindent
\textbf{Consent to publish} Consent to publish and disseminate results was received from all human participants.

\noindent
\textbf{Competing interests} The authors declare no competing interests.

\noindent
\textbf{Code availability} The code for \resides and \residem can downloaded from here \href{https://github.com/OSU-MR/reside}{https://github.com/OSU-MR/reside}

\noindent
\textbf{Authors' contributions} S. Liu coded \reside and drafted the first version of the manuscript. M. Shafique optimized \reside, compared it to VarNet, and created the final figures and tables. P. Schniter contributed by providing valuable feedback for optimizing \reside and assisted with manuscript preparation. R. Ahmad supported the project by assisting with data acquisition, troubleshooting \reside, study design, and manuscript preparation. 

\clearpage
\bibliography{root.bib}

\begin{thebibliography}{10}
\providecommand{\doi}[1]{\url{https://doi.org/#1}}
\bibcommenthead

\bibitem[\protect\citeauthoryear{Ravishankar et~al.}{2020}]{ravishankar2019image}
Ravishankar S, Ye JC, Fessler JA.
\newblock Image Reconstruction: From Sparsity to Data-Adaptive Methods and Machine Learning.
\newblock Proceedings of the IEEE. 2020;108(1):86--109.

\bibitem[\protect\citeauthoryear{Pruessmann et~al.}{1999}]{pruessmann1999sense}
Pruessmann KP, Weiger M, Scheidegger MB, Boesiger P.
\newblock {SENSE}: Sensitivity encoding for fast {MRI}.
\newblock Magnetic Resonance in Medicine. 1999;42(5):952--962.

\bibitem[\protect\citeauthoryear{Lustig et~al.}{2007}]{lustig2007sparse}
Lustig M, Donoho D, Pauly JM.
\newblock Sparse {MRI}: The application of compressed sensing for rapid {MR} imaging.
\newblock Magnetic Resonance in Medicine. 2007;58(6):1182--1195.

\bibitem[\protect\citeauthoryear{Forman et~al.}{2016}]{forman2016compressed}
Forman C, Wetzl J, Hayes C, Schmidt M.
\newblock Compressed Sensing: a Paradigm Shift in {MRI}.
\newblock MAGNETOM Flash. 2016;p.~19.

\bibitem[\protect\citeauthoryear{Zbontar et~al.}{2018}]{zbontar2018fastmri}
Zbontar J, Knoll F, Sriram A, Muckley MJ, Bruno M, Defazio A, et~al.
\newblock {fastMRI: A}n open dataset and benchmarks for accelerated {MRI}.
\newblock arXiv:181108839. 2018;.

\bibitem[\protect\citeauthoryear{Aggarwal et~al.}{2018}]{aggarwal2018modl}
Aggarwal HK, Mani MP, Jacob M.
\newblock MoDL: Model-based deep learning architecture for inverse problems.
\newblock IEEE transactions on medical imaging. 2018;38(2):394--405.

\bibitem[\protect\citeauthoryear{Sriram et~al.}{2020}]{sriram2020end}
Sriram A, Zbontar J, Murrell T, Defazio A, Zitnick CL, Yakubova N, et~al.
\newblock End-to-end variational networks for accelerated {MRI} reconstruction.
\newblock In: Medical Image Computing and Computer Assisted Intervention--MICCAI 2020: 23rd International Conference, Lima, Peru, October 4--8, 2020, Proceedings, Part II 23. Springer; 2020. p. 64--73.

\bibitem[\protect\citeauthoryear{Wang et~al.}{2021}]{wang2021deep}
Wang S, Xiao T, Liu Q, Zheng H.
\newblock Deep learning for fast MR imaging: a review for learning reconstruction from incomplete k-space data.
\newblock Biomedical Signal Processing and Control. 2021;68:102579.

\bibitem[\protect\citeauthoryear{Chen et~al.}{2020}]{chen2020ocmr}
Chen C, Liu Y, Schniter P, Tong M, Zareba K, Simonetti O, et~al.
\newblock {OCMR} (v1.0)--{O}pen-Access Multi-Coil k-Space Dataset for Cardiovascular Magnetic Resonance Imaging.
\newblock arXiv preprint arXiv:200803410. 2020;.

\bibitem[\protect\citeauthoryear{Zeng et~al.}{2021}]{zeng2021review}
Zeng G, Guo Y, Zhan J, Wang Z, Lai Z, Du X, et~al.
\newblock A review on deep learning {MRI} reconstruction without fully sampled k-space.
\newblock BMC Medical Imaging. 2021;21(1):195.

\bibitem[\protect\citeauthoryear{Ulyanov et~al.}{2018}]{ulyanov2018deep}
Ulyanov D, Vedaldi A, Lempitsky V.
\newblock Deep image prior.
\newblock In: Proceedings of the IEEE Conference on Computer Vision and Pattern Recognition; 2018. p. 9446--9454.

\bibitem[\protect\citeauthoryear{Heckel and Hand}{2018}]{heckel2018deep}
Heckel R, Hand P.
\newblock Deep decoder: Concise image representations from untrained non-convolutional networks.
\newblock arXiv preprint arXiv:181003982. 2018;.

\bibitem[\protect\citeauthoryear{Lehtinen et~al.}{2018}]{lehtinen2018noise2noise}
Lehtinen J, Munkberg J, Hasselgren J, Laine S, Karras T, Aittala M, et~al.
\newblock {Noise2Noise}: Learning image restoration without clean data.
\newblock arXiv preprint arXiv:180304189. 2018;.

\bibitem[\protect\citeauthoryear{Krull et~al.}{2019}]{krull2019noise2void}
Krull A, Buchholz TO, Jug F.
\newblock {Noise2Void}-learning denoising from single noisy images.
\newblock In: Proceedings of the IEEE/CVF Conference on Computer Vision and Pattern Recognition; 2019. p. 2129--2137.

\bibitem[\protect\citeauthoryear{Batson and Royer}{2019}]{batson2019noise2self}
Batson J, Royer L.
\newblock {Noise2Self}: Blind denoising by self-supervision.
\newblock In: International Conference on Machine Learning; 2019. p. 524--533.

\bibitem[\protect\citeauthoryear{Quan et~al.}{2020}]{quan2020self2self}
Quan Y, Chen M, Pang T, Ji H.
\newblock Self2self with dropout: Learning self-supervised denoising from single image.
\newblock In: Proceedings of the IEEE/CVF Conference on Computer Vision and Pattern Recognition; 2020. p. 1890--1898.

\bibitem[\protect\citeauthoryear{Xu et~al.}{2020}]{xu2020noisy}
Xu J, Huang Y, Cheng MM, Liu L, Zhu F, Xu Z, et~al.
\newblock {Noisy-As-Clean}: Learning self-supervised denoising from corrupted image.
\newblock IEEE Transactions on Image Processing. 2020;29:9316--9329.

\bibitem[\protect\citeauthoryear{Zhussip et~al.}{2019}]{zhussip2019extending}
Zhussip M, Soltanayev S, Chun SY.
\newblock Extending Stein's unbiased risk estimator to train deep denoisers with correlated pairs of noisy images.
\newblock Advances in Neural Information Processing Systems. 2019;32.

\bibitem[\protect\citeauthoryear{Yoo et~al.}{2021}]{yoo2021time}
Yoo J, Jin KH, Gupta H, Yerly J, Stuber M, Unser M.
\newblock Time-dependent deep image prior for dynamic {MRI}.
\newblock IEEE Transactions on Medical Imaging. 2021;40(12):3337--3348.

\bibitem[\protect\citeauthoryear{Bell et~al.}{2023}]{bell2023robust}
Bell E, Liang S, Qu Q, Ravishankar S.
\newblock Robust Self-Guided Deep Image Prior.
\newblock In: 2023 IEEE International Conference on Acoustics, Speech and Signal Processing; 2023. p. 1--5.

\bibitem[\protect\citeauthoryear{Hamilton et~al.}{2023}]{hamilton2023low}
Hamilton JI, Truesdell W, Galizia M, Burris N, Agarwal P, Seiberlich N.
\newblock A low-rank deep image prior reconstruction for free-breathing ungated spiral functional {CMR} at {0.55 T} and {1.5 T}.
\newblock Magnetic Resonance Materials in Physics, Biology and Medicine. 2023;p. 1--14.

\bibitem[\protect\citeauthoryear{Ak{\c{c}}akaya et~al.}{2019}]{akccakaya2019scan}
Ak{\c{c}}akaya M, Moeller S, Weing{\"a}rtner S, U{\u{g}}urbil K.
\newblock Scan-specific robust artificial-neural-networks for k-space interpolation ({RAKI}) reconstruction: {D}atabase-free deep learning for fast imaging.
\newblock Magnetic Resonance in Medicine. 2019;81(1):439--453.

\bibitem[\protect\citeauthoryear{Zhang et~al.}{2022}]{zhang2022residual}
Zhang C, Moeller S, Demirel OB, U{\u{g}}urbil K, Ak{\c{c}}akaya M.
\newblock Residual {RAKI}: A hybrid linear and non-linear approach for scan-specific k-space deep learning.
\newblock NeuroImage. 2022;256:119248.

\bibitem[\protect\citeauthoryear{Griswold et~al.}{2002}]{griswold2002generalized}
Griswold MA, Jakob PM, Heidemann RM, Nittka M, Jellus V, Wang J, et~al.
\newblock Generalized autocalibrating partially parallel acquisitions ({GRAPPA}).
\newblock Magnetic Resonance in Medicine. 2002;47(6):1202--1210.

\bibitem[\protect\citeauthoryear{Yaman et~al.}{2020}]{yaman2020self}
Yaman B, Hosseini SAH, Moeller S, Ellermann J, U{\u{g}}urbil K, Ak{\c{c}}akaya M.
\newblock Self-supervised learning of physics-guided reconstruction neural networks without fully sampled reference data.
\newblock Magnetic Resonance in Medicine. 2020;84(6):3172--3191.

\bibitem[\protect\citeauthoryear{Moran et~al.}{2020}]{moran2020noisier2noise}
Moran N, Schmidt D, Zhong Y, Coady P.
\newblock {Noisier2Noise}: Learning to denoise from unpaired noisy data.
\newblock In: Proceedings of the IEEE/CVF Conference on Computer Vision and Pattern Recognition; 2020. p. 12064--12072.

\bibitem[\protect\citeauthoryear{Millard and Chiew}{2023}]{millard2023theoretical}
Millard C, Chiew M.
\newblock A theoretical framework for self-supervised {MR} image reconstruction using sub-sampling via variable density {Noisier2Noise}.
\newblock IEEE Transactions on Computational Imaging. 2023;.

\bibitem[\protect\citeauthoryear{Cole et~al.}{2020}]{cole2020unsupervised}
Cole EK, Pauly JM, Vasanawala SS, Ong F.
\newblock Unsupervised {MRI} reconstruction with generative adversarial networks.
\newblock arXiv preprint arXiv:200813065. 2020;.

\bibitem[\protect\citeauthoryear{Eldar}{2009}]{eldar2009}
Eldar YC.
\newblock Generalized {SURE} for Exponential Families: Applications to Regularization.
\newblock IEEE Transactions on Signal Processing. 2009;57(2):471--481.

\bibitem[\protect\citeauthoryear{Aggarwal et~al.}{2023}]{aggarwal2022ensure}
Aggarwal HK, Pramanik A, John M, Jacob M.
\newblock ENSURE: A General Approach for Unsupervised Training of Deep Image Reconstruction Algorithms.
\newblock IEEE Transactions on Medical Imaging. 2023;42(4):1133--1144.

\bibitem[\protect\citeauthoryear{Zhussip et~al.}{2019}]{zhussip2019training}
Zhussip M, Soltanayev S, Chun SY.
\newblock Training Deep Learning Based Image Denoisers From Undersampled Measurements Without Ground Truth and Without Image Prior.
\newblock In: Proceedings of the IEEE/CVF Conference on Computer Vision and Pattern Recognition (CVPR); 2019. .

\bibitem[\protect\citeauthoryear{Liu et~al.}{2022}]{liu2022mri}
Liu S, Schniter P, Ahmad R.
\newblock {MRI} Recovery with a Self-Calibrated Denoiser.
\newblock In: 2022 IEEE International Conference on Acoustics, Speech and Signal Processing; 2022. p. 1351--1355.

\bibitem[\protect\citeauthoryear{Block et~al.}{2007}]{block2007undersampled}
Block KT, Uecker M, Frahm J.
\newblock Undersampled radial {MRI} with multiple coils. Iterative image reconstruction using a total variation constraint.
\newblock Magnetic Resonance in Medicine. 2007;57(6):1086--1098.

\bibitem[\protect\citeauthoryear{Ahmad and Schniter}{2015}]{ahmad2015iteratively}
Ahmad R, Schniter P.
\newblock Iteratively Reweighted $\ell_1$ Approaches to Sparse Composite Regularization.
\newblock IEEE Trans Comp Imag. 2015 Dec;10(2):220--235.

\bibitem[\protect\citeauthoryear{Venkatakrishnan et~al.}{2013}]{venkatakrishnan2013plug}
Venkatakrishnan SV, Bouman CA, Wohlberg B.
\newblock Plug-and-play priors for model based reconstruction.
\newblock In: 2013 IEEE Global Conference on Signal and Information Processing; 2013. p. 945--948.

\bibitem[\protect\citeauthoryear{Ono}{2017}]{ono2017primal}
Ono S.
\newblock Primal-Dual Plug-and-Play Image Restoration.
\newblock IEEE Signal Process Lett. 2017;24(8):1108--1112.

\bibitem[\protect\citeauthoryear{Ahmad et~al.}{2020}]{ahmad2020plug}
Ahmad R, Bouman CA, Buzzard GT, Chan S, Liu S, Reehorst ET, et~al.
\newblock Plug-and-play methods for magnetic resonance imaging: Using denoisers for image recovery.
\newblock IEEE Signal Processing Magazine. 2020;37(1):105--116.

\bibitem[\protect\citeauthoryear{Ravishankar and Bresler}{2010}]{ravishankar2010mr}
Ravishankar S, Bresler Y.
\newblock {MR} image reconstruction from highly undersampled k-space data by dictionary learning.
\newblock IEEE Transactions on Medical Imaging. 2010;30(5):1028--1041.

\bibitem[\protect\citeauthoryear{Wen and Chan}{2011}]{wen2011parameter}
Wen YW, Chan RH.
\newblock Parameter selection for total-variation-based image restoration using discrepancy principle.
\newblock IEEE Transactions on Image Processing. 2011;21(4):1770--1781.

\bibitem[\protect\citeauthoryear{Shastri et~al.}{2020}]{shastri2020autotuning}
Shastri SK, Ahmad R, Schniter P.
\newblock Autotuning plug-and-play algorithms for {MRI}.
\newblock In: 2020 54th Asilomar Conference on Signals, Systems, and Computers; 2020. p. 1400--1404.

\bibitem[\protect\citeauthoryear{Buehrer et~al.}{2007}]{buehrer2007array}
Buehrer M, Pruessmann KP, Boesiger P, Kozerke S.
\newblock Array compression for {MRI} with large coil arrays.
\newblock Magnetic Resonance in Medicine. 2007;57(6):1131--1139.

\bibitem[\protect\citeauthoryear{Uecker et~al.}{2014}]{uecker2014espirit}
Uecker M, Lai P, Murphy MJ, Virtue P, Elad M, Pauly JM, et~al.
\newblock {ESPIRiT}--{A}n eigenvalue approach to autocalibrating parallel {MRI}: {W}here {SENSE} meets {GRAPPA}.
\newblock Magnetic Resonance in Medicine. 2014;71(3):990--1001.

\bibitem[\protect\citeauthoryear{Darestani and Heckel}{2021}]{darestani2021accelerated}
Darestani MZ, Heckel R.
\newblock Accelerated {MRI} with un-trained neural networks.
\newblock IEEE Transactions on Computational Imaging. 2021;7:724--733.

\bibitem[\protect\citeauthoryear{Ong}{}]{Ong2023}
Ong F.
\newblock {SigPy}. {D}ownloaded on {S}eptember 1, 2023 from \url{https://github.com/mikgroup/sigpy-mri-tutorial}.
\newblock GitHub;.

\bibitem[\protect\citeauthoryear{Danielyan et~al.}{2011}]{danielyan2011bm3d}
Danielyan A, Katkovnik V, Egiazarian K.
\newblock {BM3D} frames and variational image deblurring.
\newblock IEEE Transactions on Image Processing. 2011;21(4):1715--1728.

\bibitem[\protect\citeauthoryear{Wissmann et~al.}{2014}]{wissmann2014mrxcat}
Wissmann L, Santelli C, Segars WP, Kozerke S.
\newblock {MRXCAT}: Realistic numerical phantoms for cardiovascular magnetic resonance.
\newblock Journal of Cardiovascular Magnetic Resonance. 2014;16(1):1--11.

\bibitem[\protect\citeauthoryear{Joshi et~al.}{2022}]{joshi2022technical}
Joshi M, Pruitt A, Chen C, Liu Y, Ahmad R.
\newblock Technical Report (v1.0)--Pseudo-random Cartesian Sampling for Dynamic {MRI}.
\newblock arXiv preprint arXiv:220603630. 2022;.

\bibitem[\protect\citeauthoryear{Chen et~al.}{2019}]{chen2019sparsity}
Chen C, Liu Y, Schniter P, Jin N, Craft J, Simonetti O, et~al.
\newblock Sparsity adaptive reconstruction for highly accelerated cardiac {MRI}.
\newblock Magnetic Resonance in Medicine. 2019;81(6):3875--3887.

\bibitem[\protect\citeauthoryear{Xu et~al.}{2019}]{xu2019new}
Xu P, Chen B, Xue L, Zhang J, Zhu L, Duan H.
\newblock A new {MNF}--{BM4D} denoising algorithm based on guided filtering for hyperspectral images.
\newblock ISA Transactions. 2019;92:315--324.

\bibitem[\protect\citeauthoryear{Darestani and Heckel}{}]{Darestani2022}
Darestani MZ, Heckel R.
\newblock {ConvDecoder}. {D}ownloaded on {S}eptember 10, 2022 from \url{https://github.com/MLI-lab/ConvDecoder}.
\newblock GitHub;.

\bibitem[\protect\citeauthoryear{Zhang et~al.}{2017}]{zhang2017beyond}
Zhang K, Zuo W, Chen Y, Meng D, Zhang L.
\newblock Beyond a {G}aussian denoiser: Residual learning of deep cnn for image denoising.
\newblock IEEE Transactions on Image Processing. 2017;26(7):3142--3155.

\bibitem[\protect\citeauthoryear{Niu et~al.}{2021}]{niu2021review}
Niu Z, Zhong G, Yu H.
\newblock A review on the attention mechanism of deep learning.
\newblock Neurocomputing. 2021;452:48--62.

\bibitem[\protect\citeauthoryear{Xue et~al.}{2024}]{xue2024imaging}
Xue H, Hooper S, Rehman A, Pierce I, Treibel T, Davies R, et~al.
\newblock Imaging transformer for {MRI} denoising with the {SNR} unit training: enabling generalization across field-strengths, imaging contrasts, and anatomy.
\newblock arXiv preprint arXiv:240402382. 2024;.

\end{thebibliography}
\end{document}